\documentclass[twocolumn,aps,prl,amsmath,amssymb]{revtex4-2}
\usepackage{graphicx}
\usepackage{dcolumn}
\usepackage{bm}

\usepackage{multirow}

\usepackage{amssymb}
\usepackage{amsmath}
\usepackage{graphicx}
\usepackage{xcolor}
\usepackage{braket}
\usepackage{CJK}
\usepackage{indentfirst}
\usepackage{amsmath}
\usepackage{cases}

\def\Eq#1{Eq.~(\ref{#1})}
\def\Fig#1{Fig.~\ref{#1}}
\def\avg#1{\left\langle#1\right\rangle}

\renewcommand{\(}{\left(}
\renewcommand{\)}{\right)}

\newcommand{\<}{\langle}
\renewcommand{\>}{\rangle}

\newcommand{\bea}{\begin{eqnarray}}
\newcommand{\eea}{\end{eqnarray}}

\definecolor{UkiyoRed}{RGB}{223,126,102}
\definecolor{UkiyoGreen}{RGB}{148,181,148}
\definecolor{UkiyoYellow}{RGB}{237,199,117}
\definecolor{YinshuaiBlue}{RGB}{130,143,199}
\definecolor{YinshuaiPurple}{RGB}{134,128,207}
\definecolor{YinshuaiOrange}{RGB}{255,203,148}
\usepackage{hyperref}
\hypersetup{colorlinks=true, citecolor=blue, urlcolor=blue, linkcolor=blue}

\begin{document}
\title{Imaginary-time Mpemba effect in quantum many-body systems}

\author{Wei-Xuan Chang$^{1,2}$}
\author{Shuai Yin$^{3,4}$}
\email{yinsh6@mail.sysu.edu.cn}
\author{Shi-Xin Zhang$^{1}$}
\email{shixinzhang@iphy.ac.cn}
\author{Zi-Xiang Li$^{1,2}$}
\email{zixiangli@iphy.ac.cn}

\affiliation{$^1$Beijing National Laboratory for Condensed Matter Physics \& Institute of Physics, Chinese Academy of Sciences, Beijing 100190, China}
\affiliation{$^2$University of Chinese Academy of Sciences, Beijing 100049, China}
\affiliation{$^3$Guangdong Provincial Key Laboratory of Magnetoelectric Physics and Devices, School of Physics, Sun Yat-Sen University, Guangzhou 510275, China}
\affiliation{$^4$School of Physics, Sun Yat-Sen University, Guangzhou 510275, China}

\date{\today}
\begin{abstract}
Various exotic phenomena emerge in non-equilibrium quantum many-body systems. The Mpemba effect, denoting the situation where a hot system freezes faster than the colder one, is a counterintuitive non-equilibrium phenomenon that has attracted enduring interest for more than half a century. In this Letter,  we report a novel phenomenon of the Mpemba effect in the imaginary-time relaxation dynamics in quantum many-body systems, dubbed as imaginary-time Mpemba effect (ITME). Through numerically exact quantum Monte-Carlo (QMC) simulation, we unambiguously demonstrate that in different classes of interacting quantum models, the initial states with higher energy are relaxed faster than lower-energy initial states in the process of imaginary-time relaxation. The emergence of ITME is intimately associated with the low-energy excitations in quantum many-body systems. More crucially, since imaginary-time dynamics is broadly applied in numerical simulation on the quantum many-body ground states, the discovery of ITME potentially provides a new pathway to expedite the quantum many-body computation, particularly for QMC involving the sign problem. 
\end{abstract}
\maketitle

\textcolor{YinshuaiBlue}{\it Introduction}--- In the systems far from equilibrium, many exotic phenomena emerge beyond the conventional wisdom in equilibrium physics, attracting growing attention in recent years~\cite{Polkovnikov2011rmp,Dziarmaga2010review,Rigol2016review,Mitra2018arcmp}. Among these phenomena, the Mpemba effect stands out as the particularly intriguing and counterintuitive one, characterized by the observation that a warmer system can freeze more rapidly than a cooler one \cite{Mpemba1969}. 
Despite enduring efforts over many years~\cite{Lu2017PNAS,Israel2019PRX,Bechhoefer2021NatureReview,Lasanta2017PRL,Holtzman2022CP,Takada2021PRE,Biswas2020PRE,Kumar2020Nature,Carollo2021PRL,Prados2022PRE,Hou2022PRE,Rajesh2023PRE,Roee2024arXiv}, the unified theory governing the underlying physics of Mpemba effect is still lacking. In recent years, the Mpemba effect has been generalized to the real-time dynamics in isolated quantum systems~\cite{Ares2023NC}. The quantum version of Mpemba effect has also been extensively investigated~\cite{Ares2023NC,Bertini2024PRL,Hayakawa2023PRL,Murciano_2024_z, Zhang2024arXiv,Moroder2024arXiv,Calabrese2024arXiv, Zhang2024arXivmbl,Ares2024arXiv, turkeshi2024quantum, Caceffo_2024_z}  and experimentally demonstrated on quantum simulation platforms~\cite{Zoller2024PRL,Zhang2024arXiv2}. Fathoming the classical and quantum Mpemba effect will enormously fertilize the fundamental understanding of statistic mechanics and quantum physics, and promote the frontier of research in different areas such as non-equilibrium dynamics and quantum many-body physics.   

\begin{figure}[t]
\includegraphics[width=0.4\textwidth]{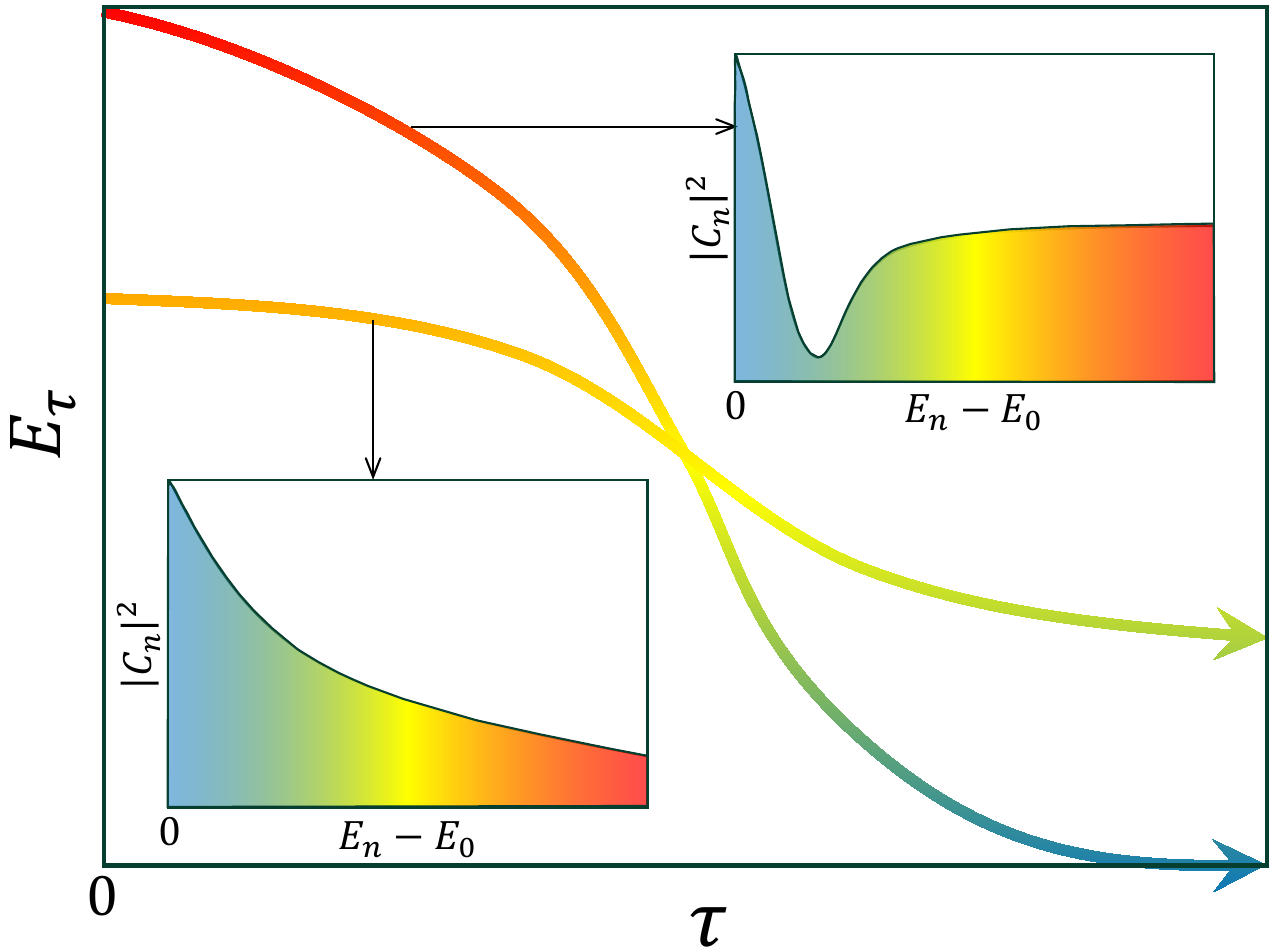}
\caption{The schematic illustration of imaginary-time Mpemba effect. The initial states with higher energy can relax faster than the lower-energy initial states under quantum imaginary-time evolution. The insets dictate the norm of the overlap between the initial state and the eigenstates of the system Hamiltonian. The higher-energy state might have a smaller overlap with low-energy eigenstates, thus exhibiting faster relaxation under imaginary-time evolution.  }
\label{Fig1}
\end{figure}

In addition to real-time dynamics, imaginary-time dynamics is also of vital importance, especially in the research of quantum many-body physics and quantum computation. As a routine tool for numerical simulation, imaginary-time evolution is generally applied to efficiently access the ground-state properties of quantum many-body systems~\cite{AssaadReview,Li2019Review,Vidal2007prl,Xiangt2008prl}. Besides, imaginary-time critical dynamics is utilized to decipher quantum critical properties both at and out of equilibrium ~\cite{Sandvik2011PRB,Yin2014prb,PolkovnikovSandvik2013prb,Yin2017PRB,Yin2022PRL,Yu2023arXiv,Zeng2024arXiv,Zeng2024arXivSUSY}. More remarkably, imaginary-time evolution has recently been proposed to approximate the ground state and thermal states on quantum computers~\cite{Motta2020naturephyscis,Nishi2021njp,Pollmann2021prxq,Yuan2019npjQI,Mao2023PRL,Zeng2022CP,Siopsis2022PRA,Fan2021,Nishi2021,Minnich2022PRXQuantum, Ding2023_z}, and imaginary-time critical dynamics has also been observed on experimental platforms of quantum computers~\cite{Zhang2023}. Owing to these inspiring progresses, a natural question is whether the celebrated Mpemba effect can emerge in the imaginary-time quantum dynamics.

\begin{figure*}[t]
\includegraphics[width=0.95\textwidth]{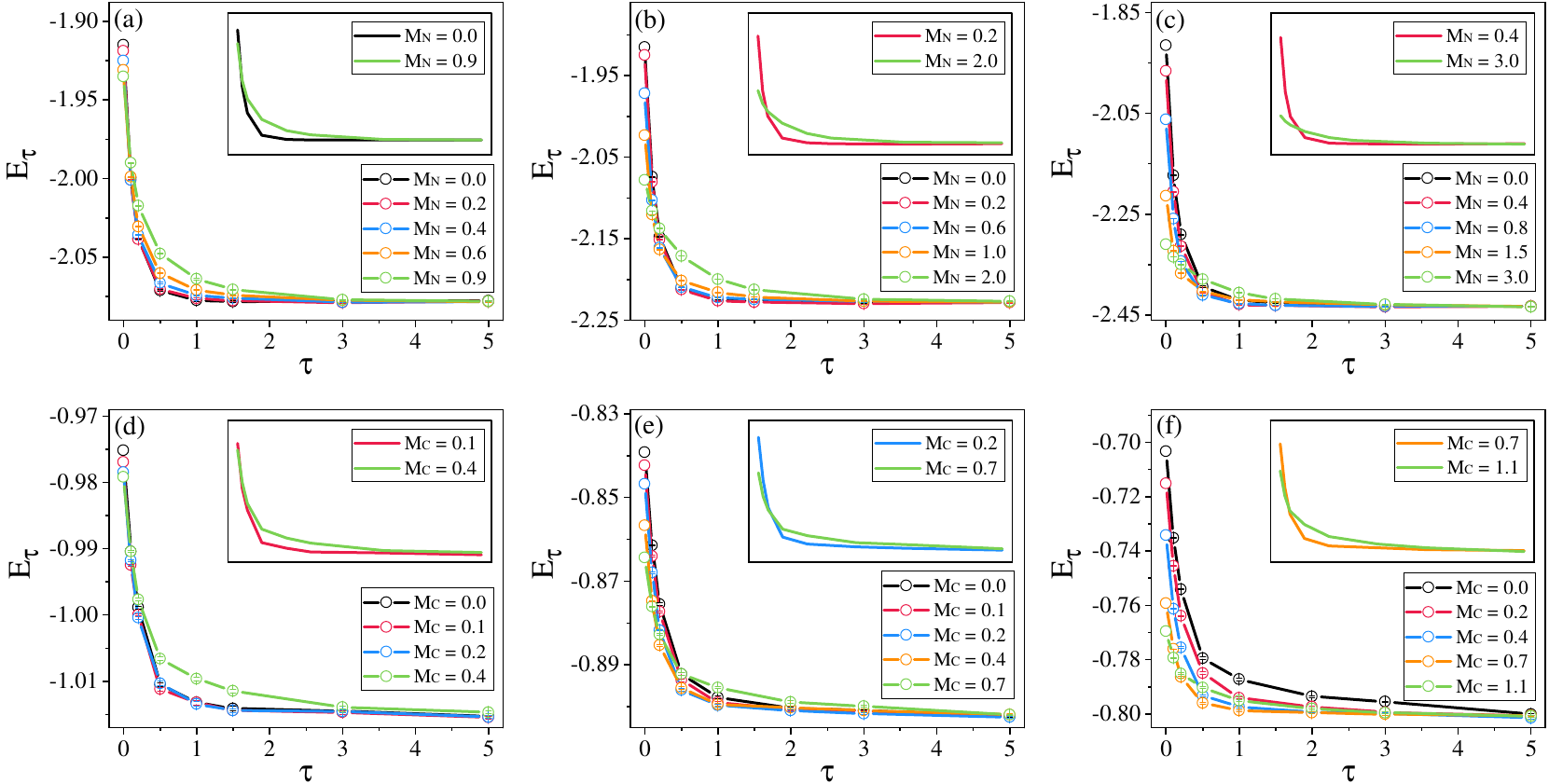}
\caption{(a)-(c): The results of energy $E_\tau$ in the imaginary-time dynamics versus imaginary time $\tau$ for $\pi$-flux square Hubbard model. The lattice size is $L\times L$ with $L=20$. The AFM mean-field initial states with different AFM order-parameter fields $M_N$ are adopted. The Hubbard interaction strength is $U=4.0$ (a), $U=5.5$ (b) and $U=7.0$ (c), respectively; (d)-(f): The results of energy $E_\tau$ versus imaginary time $\tau$ for spinless $t$-$V$ honeycomb model. The lattice size is $2\times L\times L$ with $L=12$. The CDW mean-field initial states with different CDW order-parameter fields $M_C$ are adopted. The repulsive density interaction strength is $V=1.1$ (d), $V=1.35$ (e) and $V=1.6$ (f), respectively. The insets are the results for the two initial states associated with the faster convergence and lowest initial energy, respectively. The crossing of two curves manifests the occurrence of ITME.  }
\label{Fig2}
\end{figure*}

To address this critical question, we perform numerically exact quantum Monte-Carlo (QMC) simulation to investigate the imaginary-time dynamics in various classes of interacting fermionic models. We observe that the higher-energy initial state is relaxed to the ground state faster than the lower-energy initial state under imaginary evolution, reminiscent of the Mpemba effect in real-time thermal relaxation. We dub such a novel phenomenon as the imaginary-time Mpemba effect (ITME). ITME generically occurs in different interacting models and is intimately associated with the low-energy excitations of the many-body system, typically arising from the quantum critical point (QCP) or Goldstone modes, as evidenced by our numerical results. The findings of ITME in quantum many-body systems not only enrich our knowledge of anomalous non-equilibrium dynamics 
but also have potential practical applications in improving the numerical simulation efficiency of quantum many-body problems~\cite{Li2019Review,AssaadReview}.

\textcolor{YinshuaiBlue}{\it Theoretical setup:}---
For a quantum many-body system, the imaginary-time evolution of the state under a given Hamiltonian satisfies the corresponding imaginary-time Schr\"{o}dinger equation: 
\bea
-\frac{\partial}{\partial \tau} \ket{\psi(\tau)} = (\hat{H}(\tau) - E_{\tau}) \ket{\psi(\tau)}, 
\eea
where $E_{\tau} = \bra{\psi(\tau)} \hat{H}(\tau)\ket{\psi(\tau)}$ is the energy at time $\tau$ for keeping the normalization condition of state $\ket{\psi(\tau)}~$\cite{Yuan2019npjQI}. Here, we focus on time-independent Hamiltonian $\hat{H}$. 
At a sufficiently long imaginary time, a generic initial state evolves to the ground state of $\hat{H}$, so long as the initial state is not orthogonal to the true ground state. Consequently, imaginary-time evolution is generally utilized as a theoretical tool to evaluate ground-state properties in simulation. Considering the imaginary-time evolution of a given initial state $|\psi_I\rangle$, the expectation value of an observable at imaginary-time $\tau$ is:
\bea
\langle\hat{O}\rangle_{\tau}=\frac{\langle\psi_I|e^{-\tau \hat{H}}\hat{O} e^{-\tau\hat{H}}|\psi_I\rangle}{\langle\psi_I |e^{-2 \hat{H} \tau}|\psi_I\rangle}
\label{obser}\eea
Specifically, for the case $\hat{O}=\hat{H}$, the above equation yields the energy dynamics versus evolution time: $E_\tau = \langle\hat{H}\rangle_{\tau}=\frac{\sum_n a_n^2 e^{-2E_n\tau}E_n}{\sum_n a_n^2 e^{-2E_n\tau}}$, where $E_n$ is the $n$-th eigenvalue of $\hat{H}$ and $a_n$ is expansion coefficient of $|\psi_I\rangle$ in terms of the associated eigenstate. $E_{\tau}$ monotonically decreases with $\tau$ and converges to the ground-state energy $E_{0}$  asymptotically for $\tau\rightarrow\infty$, offering an imaginary-time analog of non-equilibrium relaxation dynamics. As aforementioned, for the thermal or quantum real-time relaxation dynamics, Mpemba effects are observed~\cite{Mpemba1969,Ares2023NC}. In this study, we identify the Mpemba effect in the imaginary-time dynamics. We perform unbiased QMC simulation (see the Supplementary Materials (SM) for a detailed introduction) to evaluate the results of observable under imaginary-time evolution\cite{AssaadReview,Li2019Review}.

\textcolor{YinshuaiBlue}{\it Interacting Dirac-fermion systems:}--- We first investigate Dirac-fermion systems, with the low-energy band structure featuring Dirac fermions. At two dimensions, a prominent property of Dirac fermions is that the local interactions are irrelevant, rendering the robustness of the Dirac semimetal (DSM) phase against weak interaction. When the interaction is sufficiently strong, the Dirac fermions acquire mass and the ground state is an insulating phase~\cite{Herbut2006prl,Roy2009PRB,Li2017NCFIQCP,Assaad2018Science,Vaezi2022PRL,Assaad2017PRL,Franz2010PRB,He2018PRB,Wu2016PRB,Li2024PRLNH}. Consequently, for the typical Dirac-fermion systems, including the Hubbard model on $\pi$-flux square lattice and honeycomb lattice at half-filling, the increase of interaction strength triggers a quantum phase transition from DSM to the ordered insulating phase. Here, we systematically investigate the non-equilibrium property of imaginary-time evolution in interacting Dirac-fermion systems, and particularly, unravel the Mpemba effect for various initial states.

We first consider 
$\pi$-flux square Hubbard model, described by the Hamiltonian as follows:
\bea
H = -\sum_{\avg{ij}\sigma} t_{ij} c^\dagger_{i\sigma} c_{j\sigma} + h.c. + U \sum_i \left(\sum_{\sigma} n_{i\sigma}-1\right)^2,
\label{Ham1}
\eea
where $t_{ij}$ is the hopping amplitude on nearest-neighbor (NN) bond $ij$ and $U$ is the strength of repulsive Hubbard interaction. We fix the gauge as $t_x = t$ for the hopping at the $x$-direction and $t_y = (-1)^x t$ for the one at the $y$-direction such that the magnetic flux in each plaquette is $\pi$. At non-interacting limit, the band structures of \Eq{Ham1} features two Dirac points located at $(\pm\frac{\pi}{2},\pm\frac{\pi}{2})$. On the square lattice, the repulsive Hubbard interaction favors N\'{e}el anti-ferromagnetic (AFM) ordering tendency. With increasing interaction strength $U$, the ground state undergoes a quantum phase transition from DSM to the AFM insulating phase, occurring at $U\simeq5.5$ \cite{Assaad2015PRB}.

The Slater-determinant wave function involving MF order is a typical simplified wave function, capturing the salient feature of the desired symmetry breaking. More crucially, it is convenient to implement a Slater-determinant state as the initial state in the numerical approach based on imaginary-time evolution such as QMC~\cite{AssaadReview,Sorella1989EPL}. Since AFM ordering is dominant in \Eq{Ham1}, we adopt the AFM MF wave function as the initial state and explore the non-equilibrium process of the imaginary-time evolution under \Eq{Ham1}. The AFM MF wave function is generated as the ground state of the Hamiltonian $H_{N} = H_{0} + M_{N} \sum_i (-1)^{i_x + i_y} (\hat{n}_{i\uparrow} - \hat{n}_{i\downarrow})$, where $H_{0}$ is the non-interacting part of \Eq{Ham1}, $M_{N}$ is AFM field determining the strength of AFM order parameter in the MF wave function. We implement imaginary-time evolution on the initial state with varying $M_{\rm{N}}$. The expectation value of observable versus imaginary time $\tau$ is evaluated by \Eq{obser}. In \Fig{Fig2}, we present the results of energy $E_{\tau}$ for several values of $U$, located at different phases in the vicinity of QCP. At late times, the values of energy decay and converge to the results of the ground state, regardless of the choice of the initial state.   

\begin{figure}[t]
\includegraphics[width=0.48\textwidth]{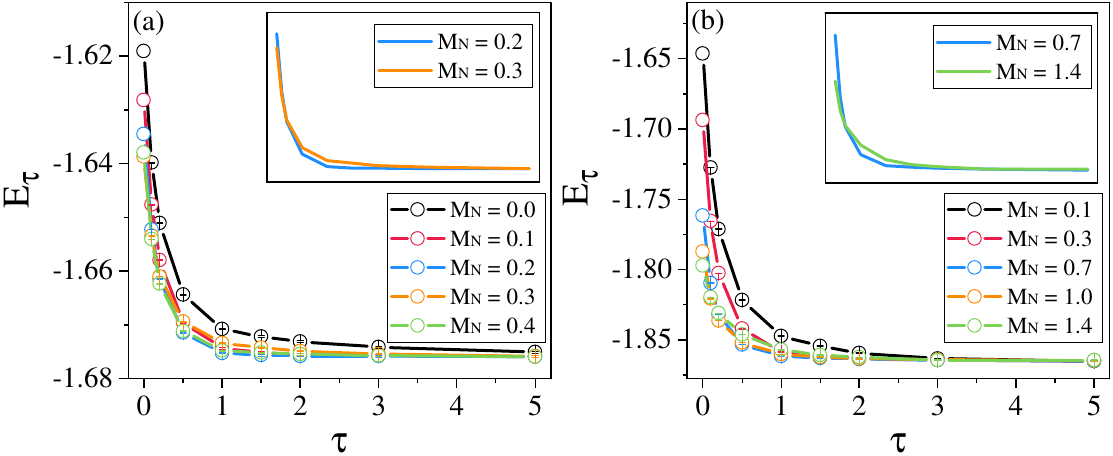}
\caption{The results of energy in the imaginary-time dynamics for Hubbard model on the square lattice. The system size is $L\times L$ with $L=20$. The AFM MF initial states with different AFM order-parameter fields $M_N$ are adopted. The Hubbard interaction strength is $U=2.0$ (a), and $U=4.0$ (b).}
\label{Fig3}
\end{figure}

We focus our study on the early-time dynamics. Intriguingly, as depicted in \Fig{Fig2}, certain states with higher energy exhibit more rapid decay of energy and converge to the ground state at an earlier imaginary time. The Mpemba effects emerge in the imaginary-time non-equilibrium process. For example, for $U=4.0$ in the DSM phase (shown in \Fig{Fig2}(a)), the state with $M_N = 0.9$ is AFM MF state with minimum energy; while the state with $M_N=0.0$, which gives higher initial energy, displays substantially rapid decay of energy and converges at shorter $\tau$. The crossing of $E_{\tau}$ curves presented in the inset is an unambiguous manifestation of the Mpemba effect. For $U=5.5$ at QCP and $U=7.0$ in the AFM phase, the pronounced ITME also appears. 
Furthermore, we employ the same procedure and investigate the imaginary-time dynamics in another typical interacting Dirac-fermion system, namely the honeycomb Hubbard model~\cite{Herbut2013prx,Sorella1992,Sorella2016prx}. The results (included in the SM) reveal the emergence of ITME as well.

To demonstrate that ITME is universal enough and arises from different types of interactions, we study a spinless Dirac-fermion model with density interaction on the honeycomb lattice, dubbed as spinless $t$-$V$ model:
\bea
 H=-t\sum_{\langle ij\rangle}c_{i}^\dagger c_{j}+V\sum_{\langle ij\rangle} \left({n_{i}-\frac{1}{2}}\right) \left({n_{j}-\frac{1}{2}}\right) \label{Ham2},
\eea
where $t$ is the NN hopping amplitude and $V$ denotes the repulsive density interaction strength. At half-filling, the ground-state phase diagram is well established by sign-free QMC simulation~\cite{Wang2014NJP,Li2015NJP,Hesselmann2016prb}. Different from Hubbard interaction, the strong density repulsion triggers a quantum phase transition from DSM to charge-density-wave (CDW) phase, occurring at $V \simeq 1.35$. Therefore, we implement CDW MF states as the initial states in the imaginary-time evolution, generated as the ground state of the MF Hamiltonian: $H_{\rm C} = H_0 + M_{\rm C}\sum_i \delta_i n_i$, where $H_0$ is the non-interacting part in \Eq{Ham2}, $M_{\rm C}$ is CDW order-parameter field, and $\delta_i =\pm 1$ if site $i$ belongs to A(B) sublattice. We evaluate the energy decay under imaginary-time evolution with varying $M_c$ in the initial states. As shown in \Fig{Fig2}, for different values of $V$ close to the QCP, the crossing of the energy curves versus $\tau$ occurs, implying the emergence of ITME. Consequently, the results of imaginary-time dynamics in different models suggest that ITME generically exists in interacting Dirac-fermion systems.

\textcolor{YinshuaiBlue}{\it Fermi surface nesting:}---Next, we proceed to investigate the imaginary-time dynamics in another important class of fermionic systems, in which the band structure possesses Fermi-surface nesting (FSN). We consider the square-lattice Hubbard model, with the associated Hamiltonian the same as \Eq{Ham1} except that magnetic flux in each plaquette is zero. At half-filling, the non-interacting band structure features the perfect FSN with momentum $(\pi,\pi)$. In stark contrast with Dirac-fermion systems, the Hubbard interactions are relevant in the presence of FSN, yielding the emergent AFM long-range ordering as long as $U>0$. 
We implement an AFM MF state, with the same definition in the simulation of the $\pi$-flux square Hubbard model, as the initial state in the imaginary-time evolution. As evidenced by the crossing point in the energy curves versus $\tau$ in \Fig{Fig3}, the results show that the Mpemba effect appears for both $U=2.0$ and $U=4.0$. Therefore, ITME also emerges on the square-lattice Hubbard model featuring FSN.

\begin{figure}[t]
\includegraphics[width=0.49\textwidth]{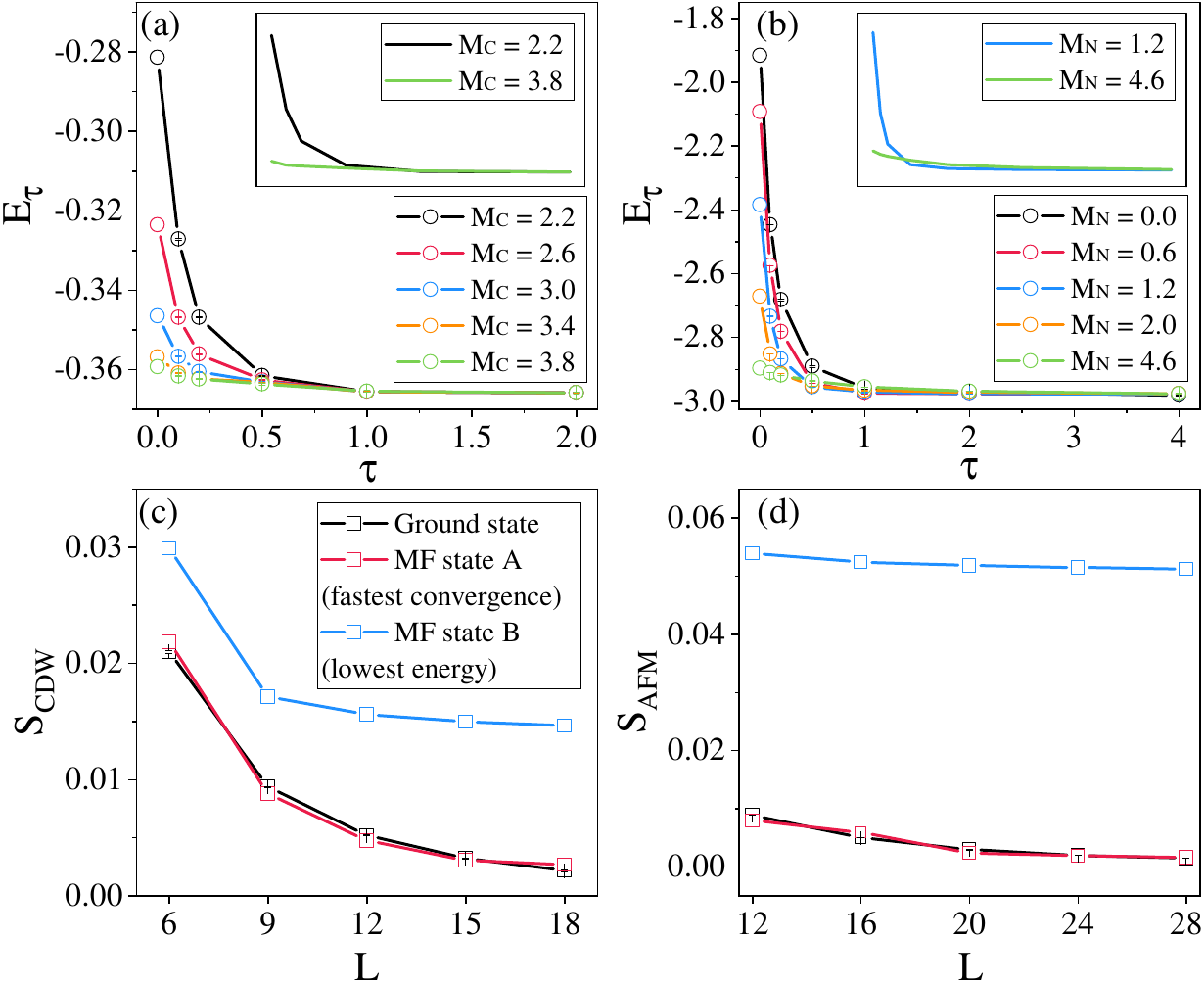}
\caption{ (a) The result of $E_{\tau}$ in spinless $t$-$V$ honeycomb model for $V=4.0$. The system size is $2\times L\times L$ with $L=12$. The CDW MF initial states with different $M_C$ are adopted. (b) The result of $E_{\tau}$ in $\pi$-flux square Hubbard model for $U=10.0$, using the AFM MF initial states with different $M_N$. The system size is $L\times L$ with $L=20$.  (c) The results of CDW structure factor $S_{\rm CDW}$ in spinless $t$-$V$ honeycomb model for $V=1.1$ in terms of different states. (d) The results of AFM structure factor $S_{\rm AFM}$ in $\pi$-flux square Hubbard model for $U=4.0$.  The red and blue squares denote the results of the MF states exhibiting the fastest convergence and the lowest energy, respectively.   }
\label{Fig4}
\end{figure}

\textcolor{YinshuaiBlue}{\it Possible underlying mechanism:}--- We numerically observe the emergence of ITME in different classes of interacting models. Although the unified theory underlying ITME is still lacking, we provide some physical understanding owing to the extensive numerical results. According to the equation $E_\tau = \langle\hat{H}\rangle_{\tau}=\frac{\sum_n a_n^2 e^{-2E_n\tau}E_n}{\sum_n a_n^2 e^{-2E_n\tau}}$, the initial state is relaxed to the ground state with short imaginary time if the overlaps of the state with the low-energy excited states of the Hamiltonian are small. ITME might emerge, if one state possesses a large overlap with the ground state and high-energy excited states, 
and the other initial state has a large overlap with low-energy excited states. The scenario is schematically described in \Fig{Fig1}. Consequently, low-energy excitation in the Hamiltonian is a prerequisite for the emergence of evident ITME. At QCP, the system is gapless due to the critical fluctuating order-parameter bosons, leading to the presence of low-energy excitation in the regime close to QCP. For all the models under consideration, the pronounced ITME emerges in the vicinity of the QCP. 
For the CDW ordered phase in the spinless $t$-$V$ model, where all the excitations are fully gapped, the results indicate the absence of evident ITME in the regime away from QCP as depicted in \Fig{Fig4} (a).  Conversely, the results of $E_{\tau}$ in the Hubbard model on a square lattice with a large value of $U$ deep in the AFM phase unambiguously show ITME (shown in \Fig{Fig4} (b)). The stark difference originates from the gapless Goldstone modes due to continuous spontaneous symmetry breaking in the AFM  phase, which is another generic scenario of stabilizing gapless excitation in quantum many-body systems. To conclude, the presence of low-energy excitation is a possible crucial ingredient for the emergence of pronounced ITME~\footnote{
Notice that the presence of low-energy excitation is only a prerequisite for the emergence of pronounced ITME. Whether the ITME occurs or not also depends on the choice of the initial states. We discuss the details of this issue in SM.}. 

 Additionally, we observe an intimate relation between ITME in early time and the ground-state observable.  In the conventional MF calculation, the MF state with minimal energy is utilized as the approximate ground state. 
 However, if ITME is present, the initial MF states with the fastest convergence speed and minimal energy are distinct. Here, we compute the observable for the two different states. For instance, we compute the square of the CDW order parameter, dubbed as structure factor $S_{\rm CDW}$ (see the definition in the SM), in the spinless $t$-$V$ model, and the results are shown in \Fig{Fig4} (c). The intriguing observation is that $S_{\rm CDW}$ for the MF state exhibiting the fastest relaxation is aligned with the true ground-state results with varying system size $L$. The alignment holds in various models including $\pi$-flux Hubbard model as shown in \Fig{Fig4} (d) (see the SM for other examples), suggesting the possible connection between the ground-state properties and early-stage imaginary-time dynamics, albeit the understanding of the correspondence is lacking and worthy of further exploration.

\textcolor{YinshuaiBlue}{\it Implication to numerical simulation:}---  As aforementioned, imaginary-time evolution is generally utilized in the quantum many-body numerical methods to access the ground-state properties. The imaginary time scale of evolution is a critical figure of merit in determining the computational hardness. Particularly, in the QMC simulation with sign problem \cite{Sugar1990PRB,Sandvik2000PRB,Troyer2005PRL,Berg2012Science,Li2019Review,Wang2015PRL,Li2016PRL,Xiang2016PRL,Li2015PRB,Hangleiter2020SA,Ringel2020PRR,Li2022arXivsign,rYAN2021PRL,Wan2022PRBsign,Xu2022PRBsign,Mondaini2022Science,Berg2023PRL,Yao2024arXivsign}, the computational time exponentially increases with the imaginary time~\cite{Troyer2005PRL}. Hence, the choice of initial state with faster relaxation speed is of paramount importance~\cite{Chang2024prb,Sorella2023PRB,zHANG2024PRR}. In the MF calculation, the optimal parameters in the MF wave function are determined by minimizing the energy. Such optimal MF wave function is widely utilized as the initial state in the conventional QMC simulation. However, regarding imaginary-time relaxation, the state with minimum initial energy is not necessarily the optimal one exhibiting the fastest convergence, owing to the existence of ITME. Therefore, the discovery and understanding of the ITME will provide a better-guiding principle for optimizing the initial states in numerical simulations, resulting in an immense improvement of the algorithm efficiency. Moreover, due to the possible connection between the emergence of ITME and ground-state properties, the imaginary-time dynamics at the early stage offer a possible avenue to unravel the ground-state properties of quantum many-body systems.

\textcolor{YinshuaiBlue}{\it Concluding remarks:}---In summary, we uncover the Mpemba effect in the imaginary-time relaxation dynamics, dubbed as ITME, in quantum many-body systems. Through numerically exact QMC simulation, we unambiguously demonstrate that the ITME emerges in a large class of interacting fermionic models. As a counterpart of the notable Mpemba effect in thermal relaxation or real-time quantum dynamics, ITME substantially enriches the understanding of non-equilibrium relaxation dynamics in quantum many-body systems. More crucially, since the imaginary-time evolution is broadly utilized in quantum many-body computation to study the ground-state properties, ITME provides a novel perspective to improve the efficiency in numerical simulations such as QMC. Notably, the discovery of ITME potentially paves a new route to mitigating the sign problem in QMC simulation by optimizing the initial state.

 Our study opens a novel direction to investigate the anomalous Mpemba effect in quantum many-body systems. Many interesting questions regarding ITME are worthy of future studies. For instance, it is intriguing to investigate the possible ITME in bosonic systems. In the present work, we perform a preliminary investigation by studying the imaginary-time dynamics in spin systems, and reveal the emergent ITME therein, as shown in the SM. The systematic study in this direction is left for future work. The unified theory of ITME is also lacking. Moreover, owing to the progress in quantum computers, it is promising to experimentally demonstrate ITME on a quantum platform.

\textcolor{YinshuaiBlue}{\it Acknowledgments}---  W. X. Chang and Z. X. Li are supported by the NSFC under Grant No. 12347107. S. Yin is supported by the National Natural Science Foundation of China (Grants No. 12075324 and No. 12222515) and the Science and Technology Projects in Guangdong Province (Grants No. 2021QN02X561). S. X. Zhang is supported by a startup grant at IOP-CAS.

\bibliography{ref}

\onecolumngrid
\newpage
\widetext
\thispagestyle{empty}

\setcounter{equation}{0}
\setcounter{figure}{0}
\setcounter{table}{0}
\renewcommand{\theequation}{S\arabic{equation}}
\renewcommand{\thefigure}{S\arabic{figure}}
\renewcommand{\thetable}{S\arabic{table}}

\pdfbookmark[0]{Supplementary Materials}{SM}
\begin{center}
    \vspace{3em}
    {\Large\textbf{Supplementary Materials for}}\\
    \vspace{1em}
    {\large\textbf{Imaginary-time quantum Mpemba effect in interacting fermionic systems}}\\
    \vspace{0.5em}
\end{center}

\subsection{Section I: Projective determinant Quantum Monte Carlo algorithm}
\label{sec:A2}
In the maintext, we perform numerically exact projective Quantum Monte Carlo (PQMC) to study the interacting fermionic models. PQMC is generally utilized to study the ground-state properties of a given quantum many-body Hamiltonian through the imaginary-time evolution. Here, we study the imaginary-time relaxation dynamics and evaluate expectation values for observable in the process of imaginary-time evolution. The expectation value under consideration at imaginary time $\tau$ is given by the following expression:
\bea
\langle \hat O \rangle_{\tau} = \frac{\langle \psi_{\tau}|\hat O|\psi_{\tau} \rangle}{\langle \psi_{\tau}|\psi_{\tau} \rangle} = \frac{\langle \psi_{I}| e^{-\tau \hat H} \hat O e^{-\tau \hat H} | \psi_{I} \rangle}{\langle \psi_{I}| e^{-2\tau \hat H} | \psi_{I} \rangle},
\label{observable}
\eea
where $| \psi_{I} \rangle$ is the initial state under imaginary-time evolution. Generally, we implement a Trotter decomposition to divide the imaginary time $\tau$ into $N_{\tau}$ imaginary-time fragments $\Delta_{\tau}$, where $\Delta_{\tau} = \tau / N_{\tau}$. Then we apply Hubbard-Stratonovich (H-S) transformation to decouple the interacting term in the Hamiltonian to a fermion-bilinear operator coupled to an auxiliary classical field so that the standard Metropolis algorithm can be employed for important sampling on the configurations of auxiliary fields \cite{AssaadReview}. In this study, we are not interested in the ground-state properties at the stage of a long time $\tau$, but instead in the early-time dynamics. Through varying imaginary time $\tau$ and calculating the expectation value of the observable at different $\tau$, we can capture the imaginary-time dynamics of the Hamiltonian. 

Here we present the details of the H-S transformation in our simulation. In our work, two typical quantum many-body systems are concerned, the spinful Hubbard model and the spinless $t$-$V$ model. In the spinful Hubbard model, the discrete H-S transformation is implemented \cite{AssaadReview}:
\bea
e^{-\frac{\Delta\tau U}{2} \(n_{i\uparrow} + n_{i\downarrow} - 1\)^2} = \sum_{l=\pm 1,\pm 2}\gamma\(l\) e^{i\sqrt{\frac{\Delta\tau U}{2}} \eta\(l\) \(n_{i\uparrow} + n_{i\downarrow} - 1\)} + \mathcal{O}\(\Delta\tau^4\),
\eea
where the auxiliary field $\eta$ and $\gamma$ are taken the following values:
\bea
& \gamma \(\pm 1\) = 1 + \sqrt{6}/3 , \\
& \gamma\(\pm 2\) = 1 - \sqrt{6}/3 ,\\
& \eta\(\pm 1\) = \pm \sqrt{2\(3-\sqrt{6}\)}, \\
& \eta\(\pm 2\) = \pm \sqrt{2\(3+\sqrt{6}\)}.
\label{HS_Hubbard}
\eea
Under the H-S transformation, the Hubbard interaction is decoupled to the fermion-bilinear operators coupled to the classical space-time dependent auxiliary field. At a given configuration of auxiliary fields, the expectation value of observable in \Eq{observable} is straightforwardly evaluated because only fermion-bilinear operators are involved in the formula after H-S transformation. We employ the standard Metropolis algorithm to implement importance sampling on the different configurations of auxiliary fields, achieving the numerical accurate results of observable as in \Eq{observable}. The numerical details of the PQMC simulation on the imaginary-time relaxation dynamics of interacting fermionic models are included in previous work \cite{Yu2023arXiv}.

In the $t$-$V$ model, the density interaction between nearest-neighbor sites is decoupled to the density channel:
\bea
e^{-\Delta_{\tau} V (n_i-\frac{1}{2})(n_j-\frac{1}{2})} = \frac{1}{2} e^{-\Delta_{\tau} V / 4} \sum_{s_{ij} = \pm 1} e^{\alpha s_{ij} (n_{i} + n_{j} - \frac{1}{2})},
\eea
where cosh $\alpha = e^{- \Delta_{\tau} V / 2}$, $s_{ij}$ is the auxiliary field defined on the bond $ij$. Notice that in this decoupling channel, the spinless $t$-$V$ model is sign-problematic, but the sign problem can be neglected in the early-time dynamics \cite{Li2015NJP,Li2022arXivsign}. 

\subsection{Section II: Imaginary-time Mpemba effect in spinful Hubbard model on honeycomb lattice}
\begin{figure*}[t]
\includegraphics[width=0.95\textwidth]{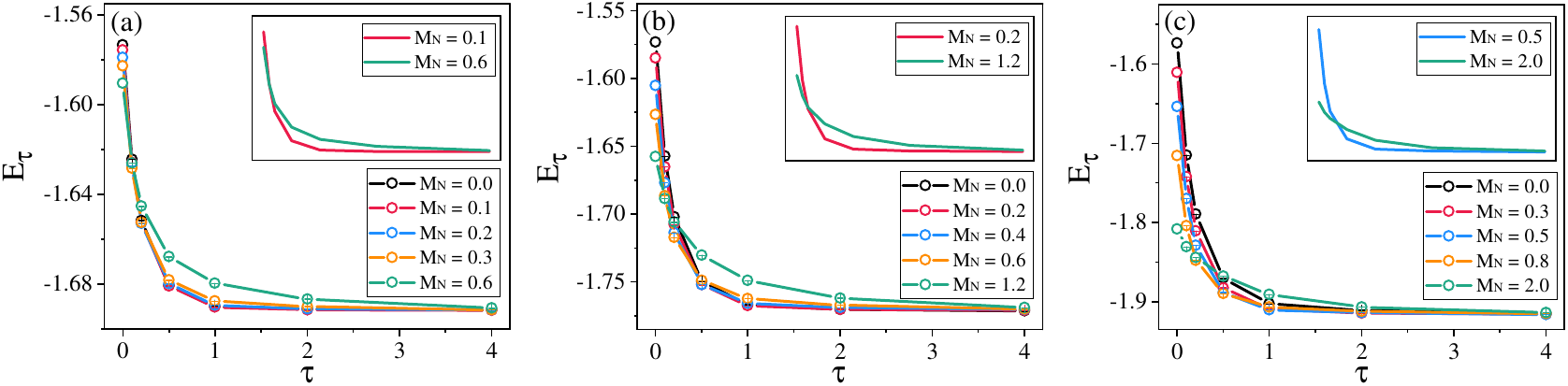}
\caption{The results of energy in the imaginary-time dynamics for the spinful honeycomb Hubbard model versus imaginary time $\tau$. The lattice size is $L\times L$ with $L=12$. The AFM mean-field initial states with different AFM order-parameter fields $M_N$ are adopted. The Hubbard interaction strength is fixed at $U=3.0$ (a), $U=3.85$ (b), and $U=5.0$ (c), respectively. The insets are the results for the two initial states associated with the faster convergence and lowest energy, respectively. The crossing point of two curves manifests the occurrence of ITME.}
\label{figS1}
\end{figure*}

In the maintext, several typical interacting fermionic systems are investigated. We demonstrate the emergence of the imaginary-time quantum Mpemba effect in a large class of fermionic quantum many-body systems. Through analysis, we find that the emergence of ITME is closely related to the low-energy excitations of quantum many-body systems. To further support our observation, we supplement our results on the spinful honeycomb Hubbard model, a well-known spinful fermion system. The corresponding Hamiltonian is described as follows:
\bea
H = -t\sum_{\avg{ij}\sigma} c^\dagger_{i\sigma} c_{j\sigma} + h.c. + U \sum_i \(n_{i\uparrow}-\frac{1}{2}\)\(n_{i\downarrow}-\frac{1}{2}\),
\label{HamS1}
\eea
where $t$ is the hopping amplitude and $U$ is the strength of repulsive Hubbard interaction. We fix the unit of energy as $t = 1$. On the honeycomb lattice, the repulsive Hubbard interaction favors N\'eel AFM ordering. With increasing interaction strength $U$, the ground state undergoes a quantum phase transition from DSM to the AFM insulating phase, and the quantum critical point occurs at $U\simeq  3.85$ \cite{Sorella2016prx}. Since the dominant order in \Eq{HamS1} is AFM order, we use the AFM MF wave function as the initial state. The AFM MF wave function is generated as the ground state of the following Hamiltonian:
\bea
H_{N} = H_{0} + M_{N} \sum_i \delta_i (\hat{n}_{i\uparrow} - \hat{n}_{i\downarrow}), 
\eea
where $H_{0}$ is the non-interacting part of \Eq{HamS1}, $M_{\rm{N}}$ is the amplitude of AFM ordering field, and $\delta_i = 1(-1)$ if site $i$ belongs to A(B) sublattice. Then, we employ the same procedure as done in the maintext and obtain energy under imaginary-time evolution for various values of Hubbard interaction strength, which are displayed in \Fig{figS1}. The crossing behavior of two curves of energy versus imaginary time $\tau$, as shown in the insets of \Fig{figS1}, unambiguously demonstrate the emergence of ITME. The results show that the ITME appears in a large regime of Hubbard interaction strength. 

\subsection{Section III: The imaginary-time relaxation dynamics of Dirac-fermion systems in the non-interacting limit}
\begin{figure}[t]
\includegraphics[width=0.6\textwidth]{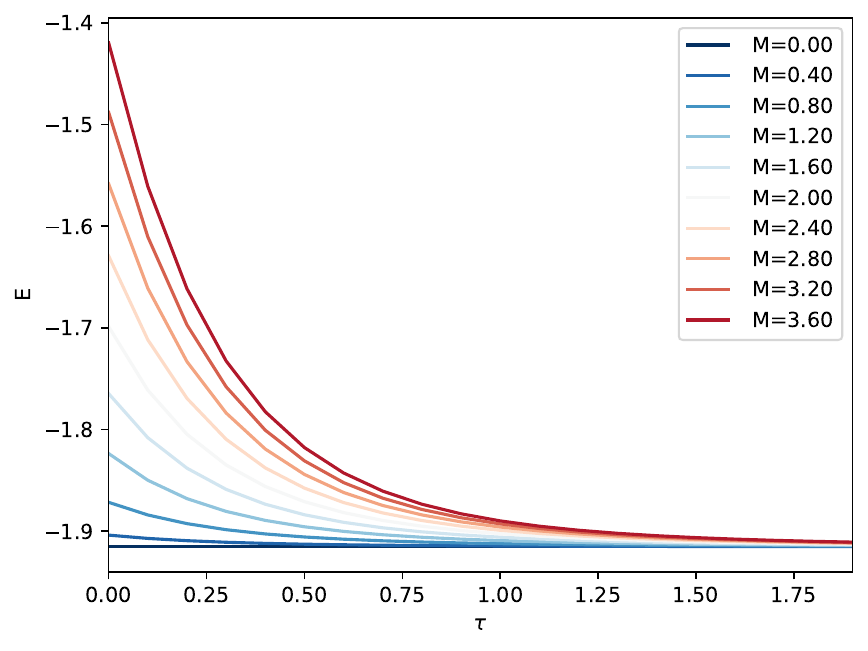}
\caption{ The results of energy versus $\tau$ in the imaginary-time relaxation for the $\pi$-flux Hubbard model on the square lattice at $U=0$. The system size is $24\times24$. The initial states are the AFM MF state with varying amplitude of ordering field $M$. The results explicitly show that with increasing $M$, the energy of the initial state increases and the convergence to the ground state is slower. The ITME is not observed in the process of imaginary-time relaxation for this class of initial states.    }
\label{dsm}
\end{figure}

In the maintext, we have systematically investigated the ITME in the strongly interacting regime. Here, we discuss Dirac-fermion systems in the non-interacting limit. We employ the same type of initial states as the simulation in the maintext, namely the MF state with certain symmetry-breaking ordering. In the non-interacting limit, the imaginary-time evolution of the observable is easily evaluated by directly solving the single-particle eigenstates of the non-interacting Hamiltonian. We study the non-interacting Dirac-fermion systems on $\pi$-flux square lattice and consider the MF state with spin AFM order. The Hamiltonian is the Eq. (3) in the maintext with fixing $U=0$. The initial states are generated as the ground state of the MF Hamiltonian $H_{N} = H_{0} + M \sum_i (-1)^{i_x + i_y} (\hat{n}_{i\uparrow} - \hat{n}_{i\downarrow})$. The numerical simulation for the free fermion dynamics is conducted using the quantum software TensorCircuit \cite{Zhang2023tensorcircuit} and the numerical results are shown in \Fig{dsm}. The results demonstrate the absence of ITME. With varying the amplitude of the MF ordering field $M$, the initial state with lower energy displays faster convergence to the ground state in the imaginary-time relaxation dynamics. We have also considered other non-interacting Dirac-fermion systems and other types of MF ordering. The numerical results are qualitatively the same and the ITME is absent in the process of imaginary-time relaxation. Additionally, for the free fermions on the square lattice without magnetic flux featuring Fermi-surface nesting, we also consider the AFM ordering which opens a gap on the Fermi surface. The ITME is also absent in the imaginary-time evolution.  

Nonetheless, although ITME is not present in the imaginary-time evolution under non-interacting Dirac-fermion Hamiltonian with the initial states of the generic MF wave function, we will demonstrate that ITME can still occur if we artificially construct certain initial states. We perform an analytical analysis of the ITME in the non-interacting Dirac-fermion systems. We consider a lattice model featuring Dirac fermions with periodic boundary conditions. Owing to the lattice translational symmetry, we can diagonalize the Hamiltonian in the space of lattice momentum. For example, for the tight-binding model involving NN hopping on the $\pi$-flux square lattice, the Hamiltonian reads: 
\bea
H_0 = -t \sum_{\avg{ij}} \delta_{\avg{ij}} c^\dagger_{i,A} c_{j,B} + h.c.,
\label{NIHam1}
\eea
 Due to the existence of $\pi$-flux in each plaquette, the unit cell on a square lattice is doubled which contains two sites. Here, $c^\dagger_{i, A(B)}$ denotes the creation operator of an electron on the site $i$ of the $A(B)$ sub-lattice, and $t$ is the amplitude of the NN hopping. We choose the gauge as $\delta_{\avg{ij}}=1$ if the bond $\avg{ij}$ is in the x-direction. $\delta_{\avg{ij}}=(-1)^{i_x+i_y}$ if the bond $\avg{ij}$ is in the y-direction and $j_y = i_y +1$.  After Fourier transform of the fermion operator by $c^\dagger_{i, A(B)} = \frac{1}{N} \sum_{\vec{k}} e^{i\vec{k}\cdot \vec{r}_i}c_{\vec{k}, A(B)}$ where $N$ is the total number of unit cell and $\vec{k}$ is the lattice momentum belonging to the first Brillouin zone,  the Hamiltonian in the lattice-momentum space is $H_0 = \sum_{\vec{k}} H_0(\vec{k}) = \sum_{\vec{k}} \phi^\dagger_{\vec{k}} h_0(\vec{k}) \phi_{\vec{k}}$, where we introduce the two-component vector $\phi_{\vec{k}} = (c_{\vec{k}, A},c_{\vec{k}, B})$. The kernel matrix $h_0(\vec{k})$ is expressed as:
\begin{equation}
\label{S7}
h_0(\vec{k})=
\begin{pmatrix}
0 & -t(e^{i\vec{\delta}_1\cdot\vec{k}}+e^{i\vec{\delta}_2\cdot\vec{k}}+e^{-i\vec{\delta}_1\cdot\vec{k}}-e^{-i\vec{\delta}_2\cdot\vec{k}}) \\
-t(e^{-i\vec{\delta}_1\cdot\vec{k}}+e^{-i\vec{\delta}_2\cdot\vec{k}}+e^{i\vec{\delta}_1\cdot\vec{k}}-e^{i\vec{\delta}_2\cdot\vec{k}}) & 0
\end{pmatrix}
\end{equation}
where $\vec{\delta}_1 = (1,0)$ and $\vec{\delta}_2 = (0,1)$ represent the two vectors of the NN bond on the square lattice. The eigenvalues of $H_0(\vec{k})$ are $(+\epsilon(\vec{k}),-\epsilon(\vec{k}))$, where $\epsilon(\vec{k}) =  2t\sqrt{\cos^2{k_x} + \sin^2{k_y}} $, and the associated single-particle eigenstates are denoted as $\alpha^\dagger_{\pm}(\vec{k})\ket{0}$. In our study, we fix the number of electrons at half-filling, namely $N_e = N$ where $N$ is the number of unit cells. The ground state of \Eq{NIHam1} is written as the many-body wave function: 
\bea
\ket{\psi_0} = \prod_{\vec{k}} \alpha^\dagger_{-}(\vec{k}) \ket{0}
, 
\eea
where $\vec{k}$ belongs to the first Brillouin zone in which the total number of lattice momentum points is $N$. 

Here, we construct two states with different energy in terms of the Hamiltonian \Eq{NIHam1}. We select the momentum $\vec{k}_1$ and replace the creation operator of the single-particle mode $\alpha_{-}^\dagger(\vec{k}_1)$ with $\tilde{\alpha}^\dagger((\vec{k}_1)= a \alpha_{-}^\dagger(\vec{k}_1) + b \alpha_{+}^\dagger(\vec{k}_1) $, with the normalized condition $a^2 + b^2 =1$. The corresponding many-body state can be written as: 
\bea
\ket{\psi_1} = \tilde{\alpha}^\dagger(\vec{k}_1) \prod_{\vec{k} \neq \vec{k}_1} \alpha^\dagger_{-}(\vec{k}) \ket{0}.
\label{psi1}
\eea
Another state is constructed by selecting a different momentum $\vec{k}_2$ and replacing the creation operator of the single-particle mode $\alpha_{-}^\dagger(\vec{k}_2)$ with $\tilde{\alpha}^\dagger((\vec{k}_2)= a \alpha_{-}^\dagger(\vec{k}_2) + b \alpha_{+}^\dagger(\vec{k}_2) $, with the normalized condition $a^2 + b^2 =1$. The corresponding many-body state can be written as: 
\bea
\ket{\psi_2} = \tilde{\alpha}^\dagger(\vec{k}_2) \prod_{\vec{k} \neq \vec{k}_2} \alpha^\dagger_{-}(\vec{k}) \ket{0}.
\label{psi2}
\eea
The eigenvalue of the single-particle state $\alpha^\dagger_{\pm}(\vec{k}_1) \ket{0}$ and $\alpha^\dagger_{\pm}(\vec{k}_2) \ket{0}$ is $\pm \epsilon_{\pm}(\vec{k}_1)$ and $\pm \epsilon_{\pm}(\vec{k}_2)$, respectively. The energy of the ground state $\ket{\psi_0}$ is $E_0 = -\sum_{\vec{k}} \epsilon_{\vec{k}}$. It is also straightforward to evaluate the energy of states $\ket{\psi_1}$ and $\ket{\psi_2}$. The corresponding results $E_1$ and $E_2$ is expressed as follows:
\bea
E_1 &=& (b^2-a^2)\epsilon(\vec{k}_1) - \sum_{\vec{k}\neq \vec{k}_1}\epsilon(\vec{k}) \\
E_2 &=& (b^2-a^2)\epsilon(\vec{k}_2) - \sum_{\vec{k}\neq \vec{k}_2}\epsilon(\vec{k}) 
\eea
Supposing that $\epsilon(\vec{k}_1)<\epsilon(\vec{k}_2)$, the energies of two initial states $\ket{\psi_1}$ and $\ket{\psi_2}$ obey $E_2 > E_1$. To investigate the relaxation dynamics of the two states under the imaginary-time evolution of $H_0$ and the possible emerged ITME, we expand the two states into the eigenstates of $H_0$. The state $\ket{\psi_1}$ possesses finite overlaps with only two eigenstates of $H_0$: the ground state $\ket{\psi_0}$ and the excited state $\ket{\tilde{\psi}_1} = \alpha^\dagger_{+}(\vec{k}_1)\prod_{\vec{k}\neq\vec{k}_1} \alpha^\dagger_{-}(\vec{k}) \ket{0}$, and $\ket{\psi_1} = a \ket{\psi_0} + b \ket{\tilde{\psi}_1}$. The state $\ket{\psi_2}$ possesses finite overlaps with only two eigenstates of $H_0$: the ground state $\ket{\psi_0}$ and the excited state $\ket{\tilde{\psi}_2} = \alpha^\dagger_{+}(\vec{k}_2)\prod_{\vec{k}\neq\vec{k}_2} \alpha^\dagger_{-}(\vec{k}) \ket{0} $, and  $\ket{\psi_2} = a \ket{\psi_0} + b \ket{\tilde{\psi}_2}$. The energy of the eigenstate $\ket{\tilde{\psi}_1}$ is $\tilde{E}_1 = E_0 + 2\epsilon(\vec{k}_1)$, and the energy of the eigenstate $\ket{\tilde{\psi}_2}$ is $\tilde{E}_2 = E_0 + 2\epsilon(\vec{k}_2)$. It is obvious that $\tilde{E}_2 > \tilde{E}_1$. Hence, according to the scenario discussed in the maintext, because $\ket{\psi_1}$ possesses larger overlap with the lower eigenstates compared with $\ket{\psi_2}$ and the overlaps of these two states with the ground state are identical, it is expected that $\ket{\psi_2}$ exhibits faster convergence in the process of imaginary-time relaxation. 

\begin{figure}[t]
\includegraphics[width=0.6\textwidth]{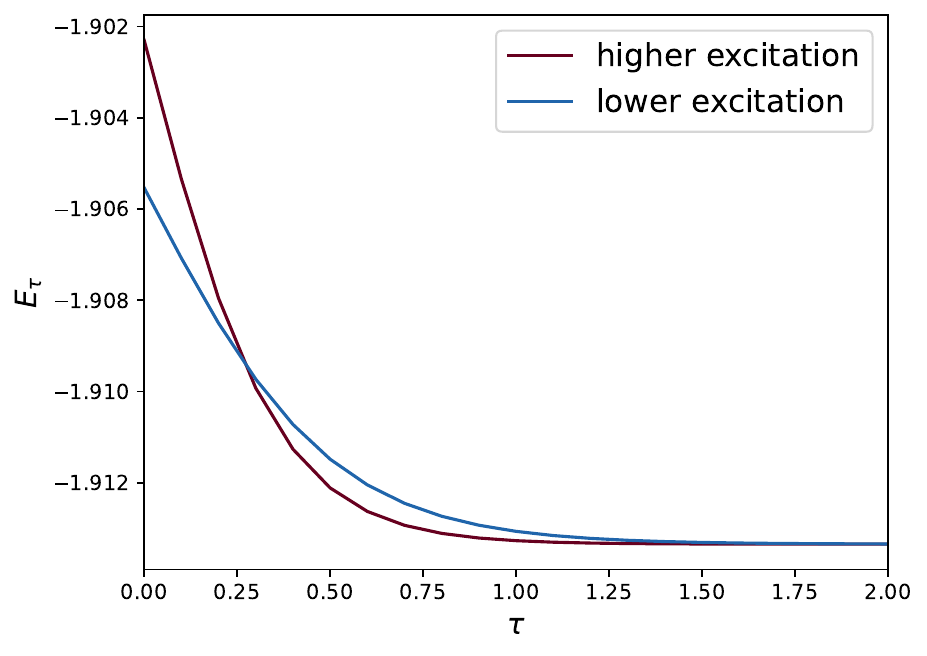}
\caption{ The results of energy versus $\tau$ in the imaginary-time relaxation for the $\pi$-flux Hubbard model on the square lattice at $U=0$. The system size is $16\times 16$. The initial states are constructed as $\ket{\psi_1}$ and $\ket{\psi_2}$ shown in \Eq{psi1} and \Eq{psi2}, respectively. The energy of initial state $\ket{\psi_2}$ is higher than the one of $\ket{\psi_1}$. However, the higher-energy state $\ket{\psi_2}$ exhibits the faster relaxation in the imaginary-tie evolution, namely the ITME emerges.    }
\label{dsm_qme}
\end{figure}

To identify the emergence of ITME for the initial states considered here in a more rigorous way, we perform numerical calculations on the energy of the two states in the process of imaginary-time evolution under the Hamiltonian of $H_0$. We perform the calculation on the $\pi$-flux square lattice, with the Hamiltonian shown in Eq. (3) of the maintext at $U=0$ and $t=1$. The initial states $\ket{\psi_1}$ and $\ket{\psi_2}$ are constructed as the procedure aforementioned, as shown in \Eq{psi1} and \Eq{psi2}. The coefficients of superposition are fixed as $a=b=1/\sqrt{2}$. We choose $\epsilon({\vec{k}_1}) = 0.005524$ and $\epsilon({\vec{k}_2})= 0.003906$ (normalized by the system size). The energy of initial state $\ket{\psi_1}$ is lower than the one of $\ket{\psi_2}$. The numerical results of the energies of the two states under imaginary-time evolution are presented in \Fig{dsm_qme}. The initial state with higher energy displays faster relaxation under imaginary-time evolution, as indicated by the crossing point of the $E_{\tau}$ curves for the two states. The evident ITME emerges in the imaginary-time evolution under $H_0$ with the initial states $\ket{\psi_1}$ and $\ket{\psi_2}$.

The above analysis and numerical results show that the emergence of ITME depends on the choice of initial states. If the evolution Hamiltonian is quadratic, we have demonstrated that the ITME is absent for generic MF initial states with varying amplitude of ordering fields. But we can artificially construct the initial states that display ITME in the process of imaginary-time evolution under the systems of free Dirac fermions. For the non-interacting Dirac-fermion systems, the pronounced ITME emerges owing to the existence of low-energy eigenstates if the initial states are constructed appropriately. 

\subsection{Section IV: The imaginary-time Mpemba effect in the Dirac semimetal phase}
\begin{figure*}[t]
\includegraphics[width=0.95\textwidth]{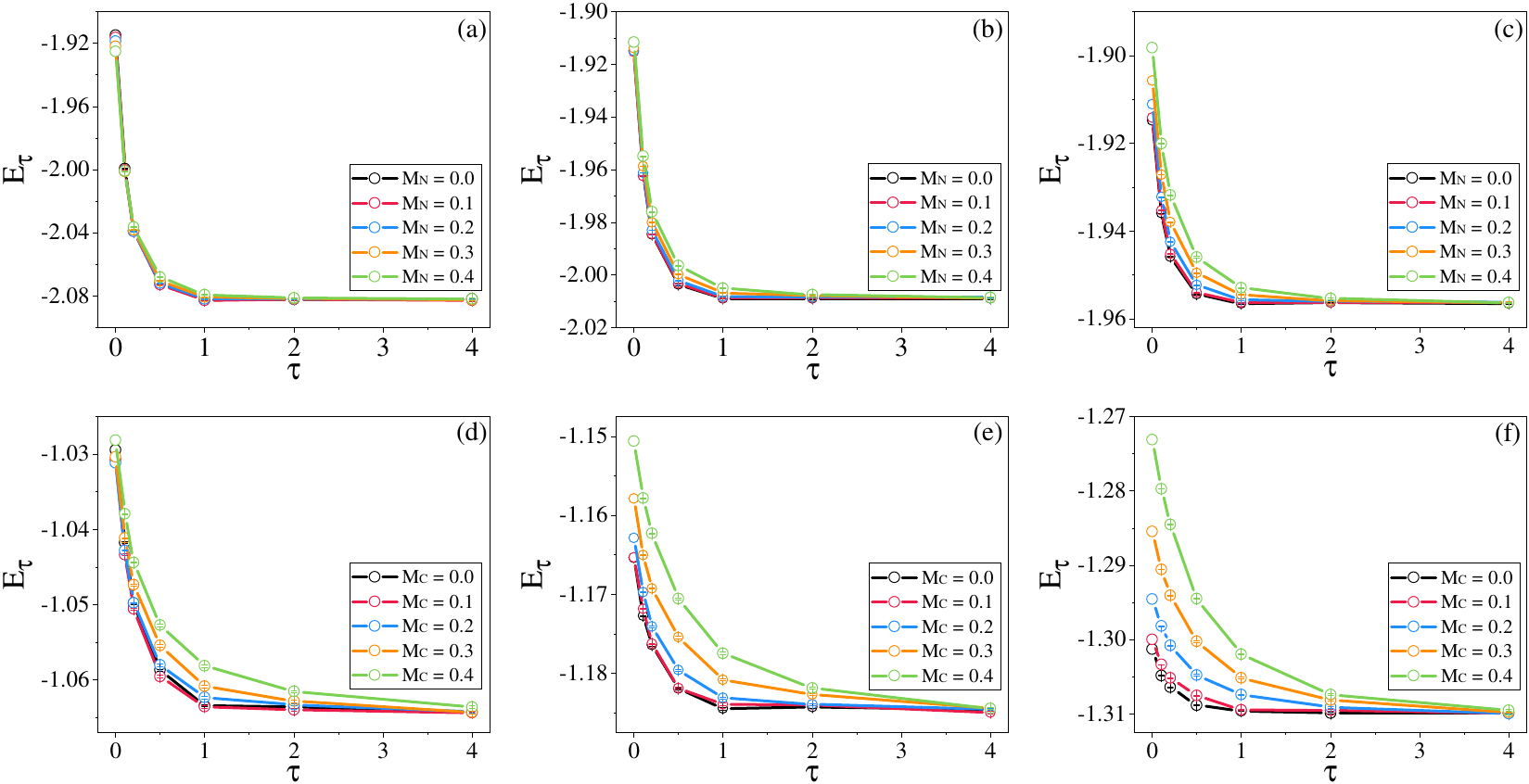}
\caption{(a)-(c): The results of energy in the imaginary-time dynamics for $\pi$-flux square Hubbard model versus imaginary time $\tau$. The lattice size is $L\times L$ with $L=20$. The AFM mean-field initial states with different AFM order-parameter fields $M_N$ are adopted. The Hubbard interaction strength is fixed at $U=4.0$ (a), $U=3.0$ (b) and $U=2.0$ (c), respectively; (d)-(f): The results of energy versus $\tau$ for spinless $t$-$V$ honeycomb model versus imaginary time $\tau$. The lattice size is $L\times L$ with $L=12$. The CDW mean-field initial states with different CDW order-parameter fields $M_C$ are adopted. The repulsive density interaction strength is fixed at $V=1.0$ (d), $V=0.75$ (e) and $V=0.5$ (f), respectively. The crossing point of two curves manifests the occurrence of ITME.}
\label{figS4}
\end{figure*}

In this section, we present the QMC results of imaginary-time relaxation in the DSM phase. In the maintext, we have shown that ITME emerges in the regime close to the QCP, with the implementation of the initial states as a class of MF states with varying amplitudes of the MF ordering field.  According to the analysis in the last section, for the interacting Dirac-fermion models considered in our study, in the non-interacting limit the ITME is absent in the imaginary-time evolution under such type of initial state. Here, we demonstrate that in the DSM phase, the pronounced ITME only exists in the vicinity of QCP. With decreasing interaction strength, the ITME gradually vanishes. We present the results of $E_{\tau}$ versus $\tau$ in the $\pi$-flux square Hubbard model and honeycomb $t$-$V$ model for different interaction strengths, as shown in \Fig{figS4}(a)-(c) and \Fig{figS4}(d)-(f). The results confirm that the pronounced ITME vanishes with decreasing interacting strength and approaching the non-interacting limit of the DSM phase. We note that the results are only applied to the MF states under investigation, other specifically designed initial states can still show ITME due to the presence of low-energy excitations in DSM phases.  

\subsection{Section V: The definitions of AFM and CDW structure factors} 
We have demonstrated the emergence of ITME in a large class of interacting fermionic models. The MF state with minimum energy is in general not the MF initial state exhibiting the fastest relaxation under imaginary-time evolution. Here, we investigate the properties of these two MF states. More explicitly, we evaluate the expectation values of physical observable in terms of the MF state with minimum energy and the one exhibiting the fastest relaxation under imaginary-time evolution, to get more insight into the emergent ITME. Here, we compute the square of order parameter, dubbed as the structure factor, which is a commonly used observable to identify the spontaneous symmetry breaking in different phases in QMC simulation. In this study, we investigate the spinful Hubbard and spinless $t$-$V$ models. In the spinful Hubbard model on honeycomb lattice and $\pi$-flux square lattice, the dominant ordering is AFM, and the corresponding AFM structure factors are defined as: 
\bea
S_{\rm{AFM}} &=& \frac{1}{2}S^{XY}_{\rm{AFM}} + S^{ZZ}_{\rm{AFM}}, \\
S^{XY}_{\rm{AFM}} &=& \frac{1}{N_s^2}\sum_{ij} \xi_i\xi_j \<(c^\dagger_{i\uparrow}c_{i\downarrow}c^\dagger_{j\downarrow}c_{j\uparrow}+c^\dagger_{i\downarrow}c_{i\uparrow}c^\dagger_{j\uparrow}c_{j\downarrow})\>, \\
S^{ZZ}_{\rm{AFM}} &=& \frac{1}{N_s^2}\sum_{ij} \xi_i\xi_j \<(c^\dagger_{i\uparrow}c_{i\uparrow}-c^\dagger_{i\downarrow}c_{i\downarrow})(c^\dagger_{j\uparrow}c_{j\uparrow}-c^\dagger_{j\downarrow}c_{j\downarrow})\>
\eea
In the spinless $t$-$V$ model on the honeycomb lattice, the repulsive density interaction favors CDW order, and the corresponding CDW structure factor is defined as:
\bea
S_{\rm{CDW}} = \frac{1}{N_s^2} \sum_{ij} \xi_i \xi_j \<(n_i-\frac{1}{2})(n_j-\frac{1}{2})\>.
\eea
Here, $i,j$ denotes the indices of the lattice site. For the AFM and CDW order, on the honeycomb lattice $\xi_i = \pm 1$ if site $i$ is on A(B) sublattice; $\xi_i = (-1)^{i_x + i_y}$ on the square lattice. The total number of lattice sites is $N_s = 2\times L_x \times L_y$ for the honeycomb lattice and $N_s =  L_x \times L_y$ got the square lattice. We compute the expectation value of the structure factors for the two MF states, namely the MF state with minimum energy and the MF state with fastest convergence under imaginary-time relaxation, and compare the results with the ground-state expectation values. We perform the calculation in the spinful Hubbard model and spinless $t$-$V$ model for different interaction strengths and system sizes.

\subsection{Section VI: Additional results of structure factors}
\begin{figure}[t]
\includegraphics[width=0.99\textwidth]{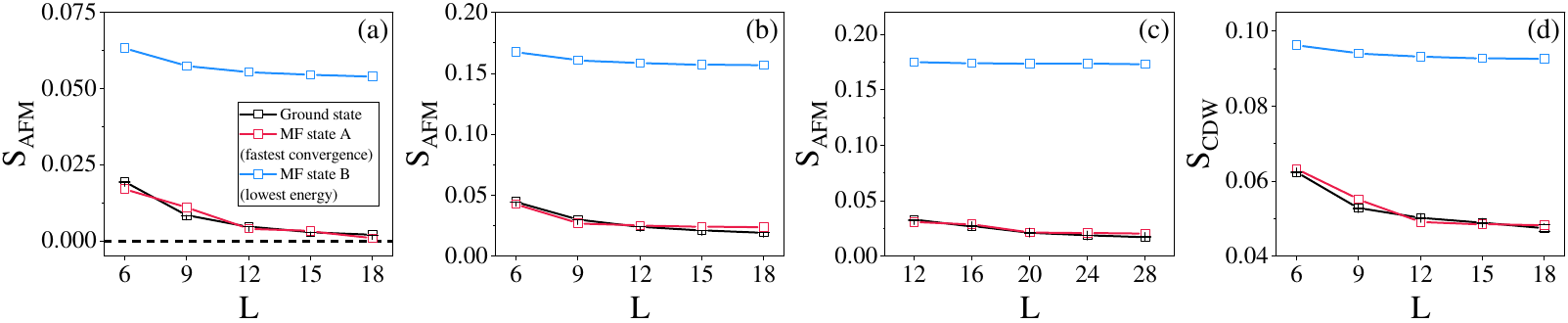}
\caption{(a) and (b) depict the results of AFM structure factor $S_{\rm AFM}$ versus system size $L$ in spinful honeycomb Hubbard model for $U=3.0$ (a) and $U=5.0$ (b) in terms of different states. The lattice size is $2\times L\times L$ with $L=12$. (c) The results of AFM structure factor $S_{\rm AFM}$ versus system size $L$ in pi-flux square Hubbard model for $U=7.0$ in terms of different states. The lattice size is $L\times L$ with $L=20$. (d) The results of CDW structure factor $S_{\rm CDW}$ versus system size $L$ in spinless honeycomb $t$-$V$ model for $V=1.6$ in terms of different states. The lattice size is $2\times L\times L$ with $L=12$. The red and blue squares denote the results of the MF wave function exhibiting the fastest convergence and the MF state with the lowest energy, respectively. }
\label{figS2}
\end{figure}

In the \Fig{Fig4}(c) and \Fig{Fig4}(d) of the maintext, we demonstrate that the AFM structure factors of the MF wave function exhibiting the fastest convergence and the ground state are closely related in the spinful $\pi$-flux square Hubbard model and spinless honeycomb $t$-$V$ model. The structure factors exhibit the same tendency with increasing system sizes. The structure factors of the fastest-converging MF state and the ground state are almost identical. To further explore this relationship, we calculate the AFM structure factor as a function of system size in the spinful honeycomb Hubbard model. The corresponding results are depicted in \Fig{figS2}(a) and \Fig{figS2}(b). We also include the structure factors for the spinful $\pi$-flux square Hubbard model and the spinless honeycomb $t$-$V$ model of other interaction strengths in \Fig{figS2}(c) and \Fig{figS2}(d), respectively. As imagined, we arrive at the same conclusions as the maintext.

\subsection{Section VII: Imaginary-time quantum Mpemba effect in one-dimensional spin models}
\begin{figure*}[t]
\includegraphics[width=0.95\textwidth]{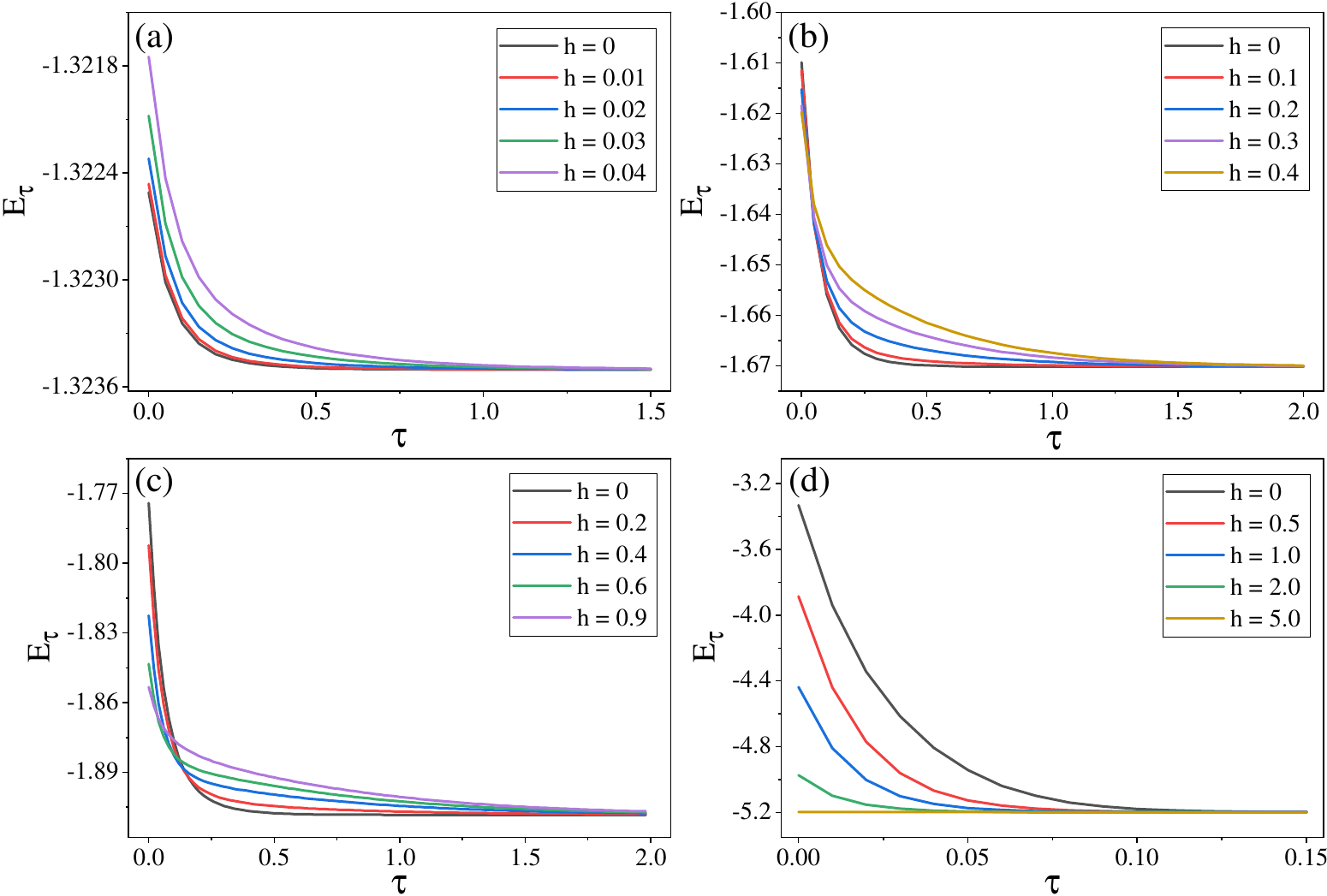}
\caption{The results of energy in the imaginary-time dynamics for 1D Heisenberg model versus imaginary time $\tau$. The lattice size is $L=16$. The AFM mean-field initial states with different AFM order-parameter fields $h$ are adopted. The $z$-component spin interaction strength is fixed at $J_z=0.1$ (a), $J_z=0.8$ (b), $J_z=1.2$ (c), and $J_z=5.0$ (d), respectively. The crossing point of the curves manifests the occurrence of ITME.}
\label{figS3}
\end{figure*}

In the maintext, we focus our study on the fermionic models. In this section, we discuss the ITME in bosonic systems. Here, we present the preliminary numerical results of the emergence of ITME in the quantum spin model. We consider the 1D XXZ model as described by the following Hamiltonian:
\bea
\hat{H} = 
J_{x}\sum_{\<i,j\>}\sigma_i^x\sigma_j^x + 
J_{y}\sum_{\<i,j\>}\sigma_i^y\sigma_j^y + 
J_{z}\sum_{\<i,j\>}\sigma_i^z\sigma_j^z
\label{HamS2}
\eea
where $J_\alpha$ is the spin interaction strength in $\alpha$-component, $\sigma_i^\alpha$ is the Pauli operator in $\alpha$-component on site $i$, and $\alpha=x,y,z$. $\<i,j\>$ denotes the nearest-neighbour sites $i$ and $j$. We fix $J_x=J_y=1$ and focus on the parameter regime $J_z >0$. The ground-state properties of \Eq{HamS2} are extensively studied by analytical and numerical approaches. With increasing $J_z$, a quantum phase transition from the paramagnetic phase and AFM ordered phase occurs at $J_z=1$. When $J_z <1$, the ground state is a Luttinger liquid, while when $J_z >1$, the ground state is an Ising AFM phase with a finite gap of excitation. Because the dominant ordering tendency is AFM in the regime $J_z >0$, we use the AFM MF wave function as the initial state. The AFM MF wave function is obtained from the ground state of the following Hamiltonian:
\bea
\hat{H}_N = 
J_{x}\sum_{\<i,j\>}\sigma_i^x\sigma_j^x + 
J_{y}\sum_{\<i,j\>}\sigma_i^y\sigma_j^y + 
h\sum_{i}(-1)^i\sigma_i^z
\eea
where $h$ is the amplitude of the AFM ordering field. We perform the exact-diagonalization (ED) calculation to study the imaginary-time relaxation dynamics of the model in \Eq{HamS2} for the AFM MF initial wave function with various $h$. Through the ED calculation, we can access the expected value of energy under imaginary-time evolution as depicted in \Eq{observable}, the results of which for various $J_z$ are shown in \Fig{figS3}. In the regime in the vicinity of the quantum critical point of $J_z=1$, the pronounced ITME appears as indicated by the crossing behavior of $E_{\tau}$ versus $\tau$, which suggests that ITME is a more universal phenomenon that exists in not only fermionic systems but also bosonic systems. Moreover, the emergence of ITME in the regime close to the QCP is consistent with the scenario that ITME is associated with the existence of low-energy excitation as discussed in the maintext.

\end{document}